\documentclass[a4paper,english]{lipics-v2019}
\usepackage{nth}
\usepackage{microtype}
\usepackage{breqn}
\usepackage{ifthen}
\usepackage{placeins}
\usepackage{complexity}
\usepackage{nicefrac}
\usepackage{todonotes}
\usepackage{siunitx}
\usepackage{amssymb}
\usepackage{color}
\usepackage{gensymb}
\usepackage[noabbrev,capitalise]{cleveref}
\usepackage{mathtools}

\hideLIPIcs
\nolinenumbers

\newcommand{\old}[1]{{}}

\title{Minimum Non-Obtuse Triangulations:\\ The CG:SHOP Challenge 2025}
\titlerunning{Minimum Non-Obtuse Triangulations: CG Challenge 2025}

\author{Sándor P.~Fekete}{Department of Computer Science, TU Braunschweig, Germany}{s.fekete@tu-bs.de}{https://orcid.org/0000-0002-9062-4241}{}
\author{Phillip Keldenich}{Department of Computer Science, TU Braunschweig, Germany}{p.keldenich@tu-bs.de}{https://orcid.org/0000-0002-6677-5090}{}
\author{Dominik Krupke}{Department of Computer Science, TU Braunschweig, Germany}{d.krupke@tu-bs.de}{https://orcid.org/0000-0003-1573-3496}{}
\author{Stefan Schirra}{Department for Simulation and Graphics, OvGU Magdeburg, Germany}{stschirr@isg.cs.uni-magdeburg.de}{https://orcid.org/0009-0006-5928-1494}{}
\authorrunning{S.~P.~Fekete, P.~Keldenich, D.~Krupke, S.~Schirra}
\Copyright{S.~P.~Fekete, P.~Keldenich, D.~Krupke, S.~Schirra}

\ccsdesc{Theory of computation $\rightarrow$ Computational geometry}
\ccsdesc{Theory of computation $\rightarrow$ Design and analysis of algorithms}

\keywords{Computational Geometry, geometric optimization, triangulation, Algorithm Engineering, contest}

\supplement{}

\funding{Work at TU Braunschweig was partially supported by the German Research Foundation (DFG), 
project ``Computational Geometry: Solving Hard Optimization Problems'' (CG:SHOP), grant FE407/21-1.} 

\acknowledgements{We thank the members of the CG:SHOP Challenge Advisory Board for their valuable input: 
William J.~Cook, Andreas Fabri, Dan Halperin, Michael Kerber, Philipp Kindermann, Joe Mitchell, Kevin Verbeek.}

\EventAcronym{CG:SHOP 2025}

\begin{document}
\maketitle
\begin{abstract}
We give an overview of the 2025 Computational Geometry Challenge
targeting the problem \textsc{}Minimum Non-Obtuse Triangulation:
Given a planar straight-line graph $G$ in the plane, defined by 
a set of points in the plane (representing vertices)
and a set of non-crossing line segments connecting them
(representing edges); the objective is to find a feasible 
non-obtuse triangulation that uses a minimum number of Steiner points. 
If no triangulation without obtuse triangles is found, the secondary objective
is to minimize the number of obtuse triangles in the triangulation.
\end{abstract}

\section{Introduction}
The ``CG:SHOP Challenge'' (Computational Geometry: Solving Hard
Optimization Problems) originated as a workshop at the 2019
Computational Geometry Week (CG Week) in Portland, Oregon in June,
2019.  The goal was to conduct a computational challenge competition
that focused attention on a specific hard geometric optimization
problem, encouraging researchers to devise and implement solution
methods that could be compared scientifically based on how well they
performed on a database of carefully selected and varied instances.
While much of computational
geometry research is theoretical, often seeking provable approximation
algorithms for \NP-hard optimization problems,
the goal of the Challenge was to set the metric of success based on
computational results on a specific set of benchmark geometric
instances. The 2019 Challenge~\cite{CGChallenge2019_JEA} focused on the problem of computing
simple polygons of minimum and maximum area for given sets of vertices in the
plane. It generated a strong response from many research
groups~\cite{area-crombez,area-tau,area-salzburg,area-exact,area-campinas,area-omega} from both the computational geometry and the combinatorial
optimization communities, and resulted in a lively exchange of
solution ideas.

Subsequently, the CG:SHOP Challenge became an event within the CG Week
program, with top performing solutions reported in the Symposium on
Computational Geometry (SoCG) proceedings. 
The schedule for the Challenge was
advanced earlier, to give an opportunity for more participation, particularly
among students, e.g., as part of course projects. 
For CG Weeks 2020, 2021, 2022, 2023, 2024 
the Challenge problems were \textsc{Minimum Convex Partition}~\cite{CGChallenge2020,SoCG2020_1,SoCG2020_2,SoCG2020_3}, 
\textsc{Coordinated Motion Planning}~\cite{CGChallenge2021,Challenge2021_1,Challenge2021_2,Challenge2021_3,SoCG2021_1,SoCG2021_2,SoCG2021_3},
\textsc{Minimum Partition into Plane Subgraphs}~\cite{CGChallenge2022_JEA,Challenge2022_1,Challenge2022_2,SoCG2022_1,SoCG2022_2,SoCG2022_3,SoCG2022_4},
\textsc{Minimum Convex Covering}~\cite{fekete2023minimum,Challenge2023_1,Challenge2023_2},
and \textsc{Maximum Polygon Packing}~\cite{fekete2024maximum,Challenge2024_1,Challenge2024_2,Challenge2024_3,Challenge2024_4},
respectively.  

The seventh edition of the Challenge in 2025 continued
this format, leading to contributions in the SoCG proceedings.

\section{The Challenge: Minimum Non-Obtuse Triangulations}

A suitable contest problem has a number of desirable properties.

\begin{enumerate}
\item The problem is of geometric nature.
\item The problem is of general scientific interest and has received previous attention.
\item Optimization problems tend to be more suitable than feasibility problems; in principle, 
  feasibility problems are also possible, but they need to be suitable for sufficiently
  fine-grained scoring to produce an interesting contest.
\item Computing optimal solutions is difficult for instances of reasonable size.
\item This difficulty is of a fundamental algorithmic nature, and not only due to
 issues of encoding or access to sophisticated software or hardware.
\item Verifying feasibility of provided solutions is relatively easy.
\end{enumerate}

In this seventh year, a call for suitable problems was communicated in May
2024. In response, a number of interesting problems were proposed for the 2025
Challenge. These were evaluated with respect to difficulty, distinctiveness
from previous years, and existing literature and related work. In the end, the
Advisory Board selected the chosen problem. Special thanks go to Mikkel Abrahamsen
(University of Copenhagen) who suggested this year's problem, motivated by a rich history
in geometry and optimization, including~\cite{LTWYC1990packing} and a wide range of
previous work described further down.

\subsection{The Problem}
The specific problem that formed the basis of the 2025 CG Challenge was the following. 

\medskip
\noindent
\textbf{Problem:} \textsc{Minimum Non-Obtuse Triangulation}\\
\textbf{Given:} A planar straight-line graph $G$, given by a set of $n$ points $P=\{p_1,\ldots,p_n\}$
in the plane and a collection $E$ of non-crossing line segments,
each connecting two points $p_i$ and $p_j$.\\
\textbf{Goal:} Inserting a suitable set $S$ of Steiner points as desired, find a 
triangulation of the convex hull of $P$ that uses $P\cup S$ as vertices, contains $E$,
such that all created triangles are non-obtuse. Among all such triangulations, find
one that minimizes the number of Steiner points.
\medskip

To make this problem more accessible, especially for students, we modified it slightly
to make it easier to find any feasible solution at all, and to have some more tractable
but non-trivial solutions available, and, thus, a more continuous difficulty spectrum.
First, we changed the problem to accept obtuse triangles, splitting the objective into
first minimizing the number of obtuse triangles, and then minimizing the number of Steiner points.
This makes any triangulation feasible and serves as a starting point for iterative improvement.
Second, we introduced a boundary to the triangulation area, which encapsulates the area of interest
into a simple polygon. While this may first sound like a complicating constraint, it actually
simplifies the problem, as Steiner points placed on the boundary are less of a risk to create further obtuse triangles that propagate through the whole instance.

\subsection{Scoring}

The scoring function was designed to reflect both solution quality and feasibility, while ensuring that even partial progress was rewarded.
Each instance was scored individually, and team scores were computed as the sum over all instances.
The scoring distinguished between solutions with and without obtuse triangles:

\begin{enumerate}
	\item \textbf{Solutions without obtuse triangles:}
	Any solution without obtuse triangles received a minimum score of $0.5$ for the respective instance.
	The best solution found by any participant, i.e., the one using the fewest Steiner points, was awarded the maximum score of $1.0$.
	Let $k_B$ be the number of Steiner points in such a best known solution and $k_Y$ be the number used by another team's solution.
	That solution received a score of:
		\(
		\frac{1}{2} + \frac{k_B}{2k_Y},
		\)
	truncated at $1.0$ if $k_Y = k_B$.
	\item \textbf{Solutions with obtuse triangles:}
		Solutions with obtuse triangles started from a base score of $0.5$.
		The score was reduced exponentially based on the number of obtuse triangles.
		Specifically, for a solution with $v$ obtuse triangles, the score was:
		\(
		\frac{1}{2} \cdot 0.97^v.
		\)
		The number of Steiner points was not considered for infeasible solutions.
\end{enumerate}

This scheme provided a smooth incentive structure: teams could improve their scores either by reducing the number of obtuse triangles or, once feasibility was reached, by minimizing Steiner points.

\subsection{Related Work}
Historically, triangulations have been crucial for applications ranging from
computer graphics and mesh generation to geographic information systems (GIS)
and finite element analysis. In computer graphics, triangulating surfaces
enables the rendering of three-dimensional objects with high precision and
efficiency. In mesh generation, triangulations help in discretizing a
continuous geometric space into simpler elements that can be easily processed
for simulations and analyses, particularly in engineering and physics.

The importance of non-obtuse triangulations in computational geometry is
particularly evident in solving partial differential equations using the finite
element method. In this method, the domain is divided into triangles, with
non-obtuse triangulations ensuring that the resulting matrices are
Stieltjes—symmetric positive definite with non-positive off-diagonal entries.
This specific matrix structure enhances the performance of iterative methods,
such as block Gauss-Seidel, improving their convergence rates. The geometric
constraint of avoiding obtuse angles in triangulations not only stabilizes
numerical calculations but also optimizes the efficiency of computational
processes, making non-obtuse triangulations a crucial consideration in
scientific computing and numerical analysis.  

Minimum Non-Obtuse Triangulations have been studied for quite a while, with
references in DCG and SoCG going back to the 1980s~\cite{baker1988nonobtuse}, and connections to
finite element analysis. Subsequent work considered bounds on the necessary
numbers of Steiner points~\cite{bern1991polynomial,bern1994linear,bern1995linear}.  More recent work~\cite{bishop2016nonobtuse} 
considers an algorithm for Planar Straight Line Graphs (PSLGs). 
The computational complexity for simple polygons and polygons with holes
remains an open question in the field.

\subsection{Instances}

The generation of appropriate instances is critical for any challenge.
Instances that are too easily solved undermine the complexity of the problem, rendering it trivial.
Conversely, instances that require significant computational resources for preprocessing or for finding viable solutions may offer an unfair advantage to teams with superior computing facilities.
This issue is amplified when the instance set is so large that it exceeds the management capacity when using a limited number of computers.

\begin{figure}
	\centering
	\begin{subfigure}[b]{0.45\textwidth}
		\centering
		\includegraphics[width=\textwidth]{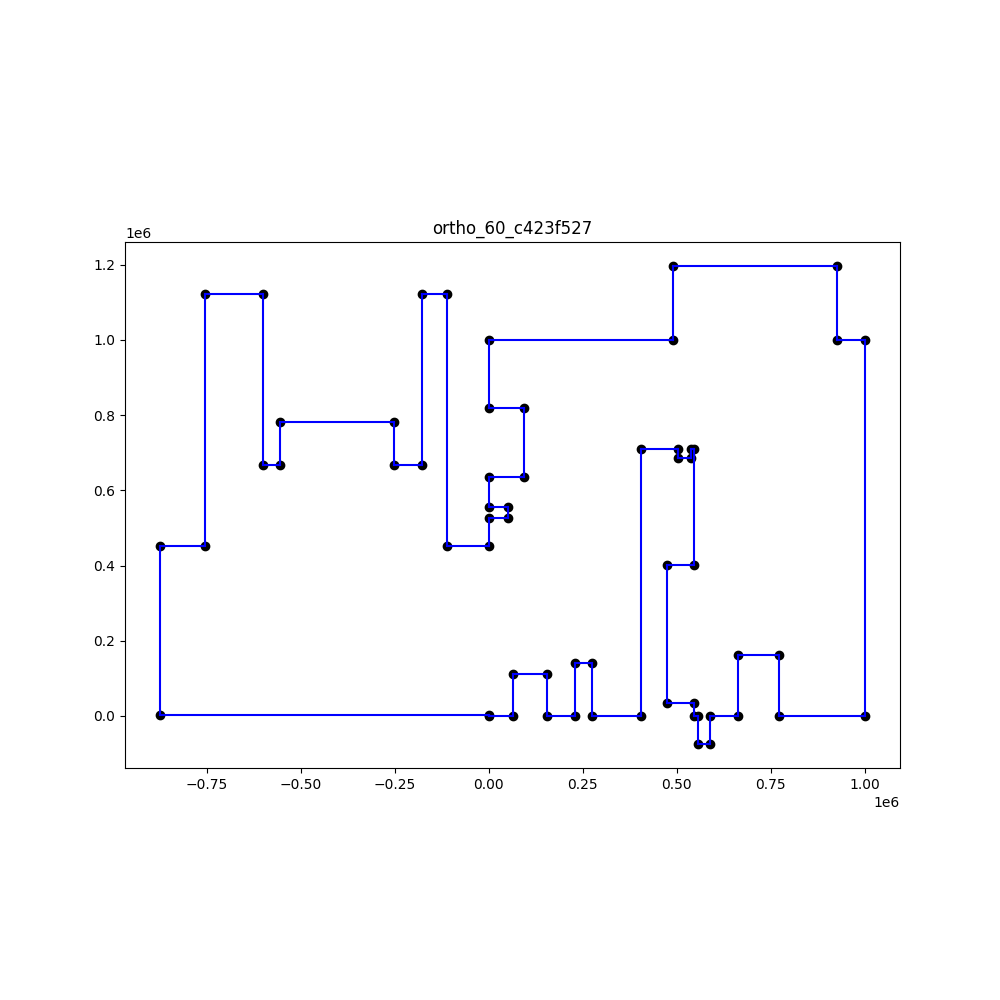}
		\caption{\texttt{ortho\_60\_c423f527}}
		\label{fig:instances:ortho}
	\end{subfigure}
	\hfill
	\begin{subfigure}[b]{0.45\textwidth}
		\centering
		\includegraphics[width=\textwidth]{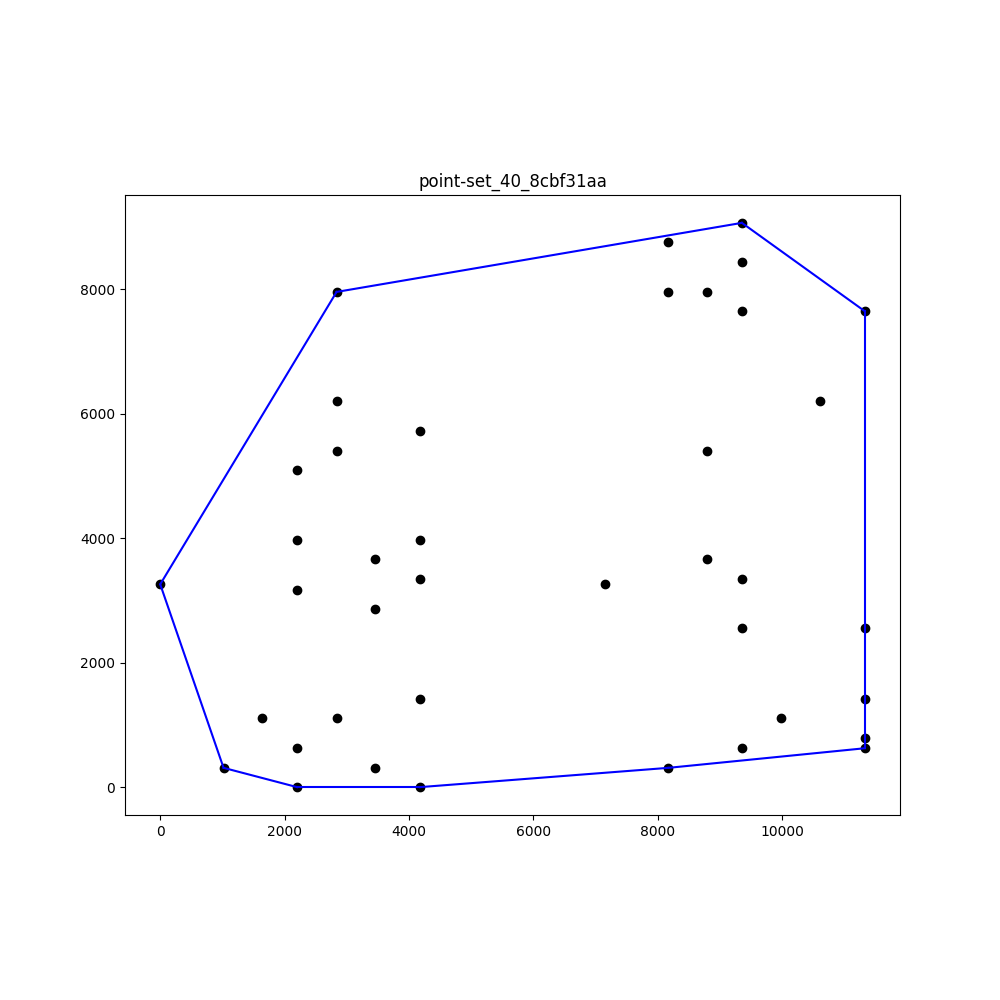}
		\caption{\texttt{point-set\_40\_8cbf31aa}}
		\label{fig:instances:point-set}
	\end{subfigure}
	\hfill
	\begin{subfigure}[b]{0.45\textwidth}
		\centering
		\includegraphics[width=\textwidth]{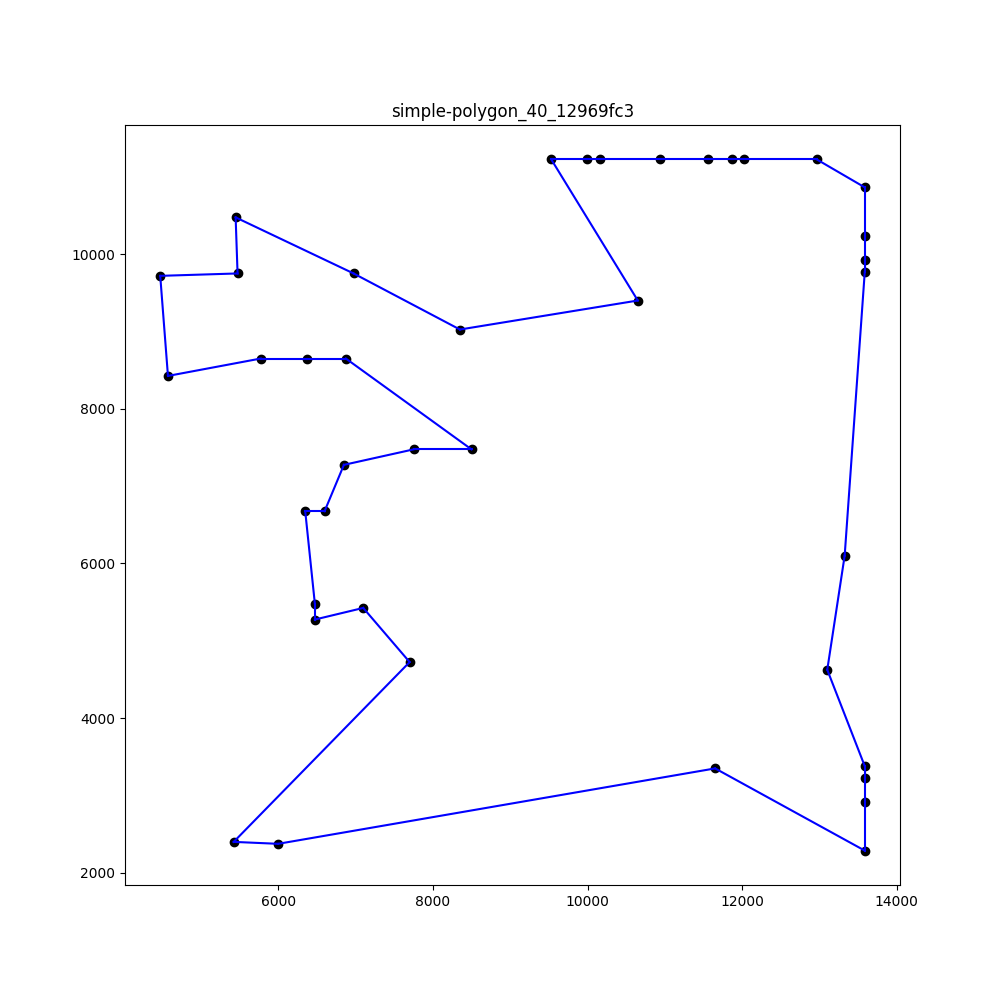}
		\caption{\texttt{simple-polygon\_40\_12969fc3}}
		\label{fig:instances:simple-polygon}
	\end{subfigure}
	\hfill
	\begin{subfigure}[b]{0.45\textwidth}
		\centering
		\includegraphics[width=\textwidth]{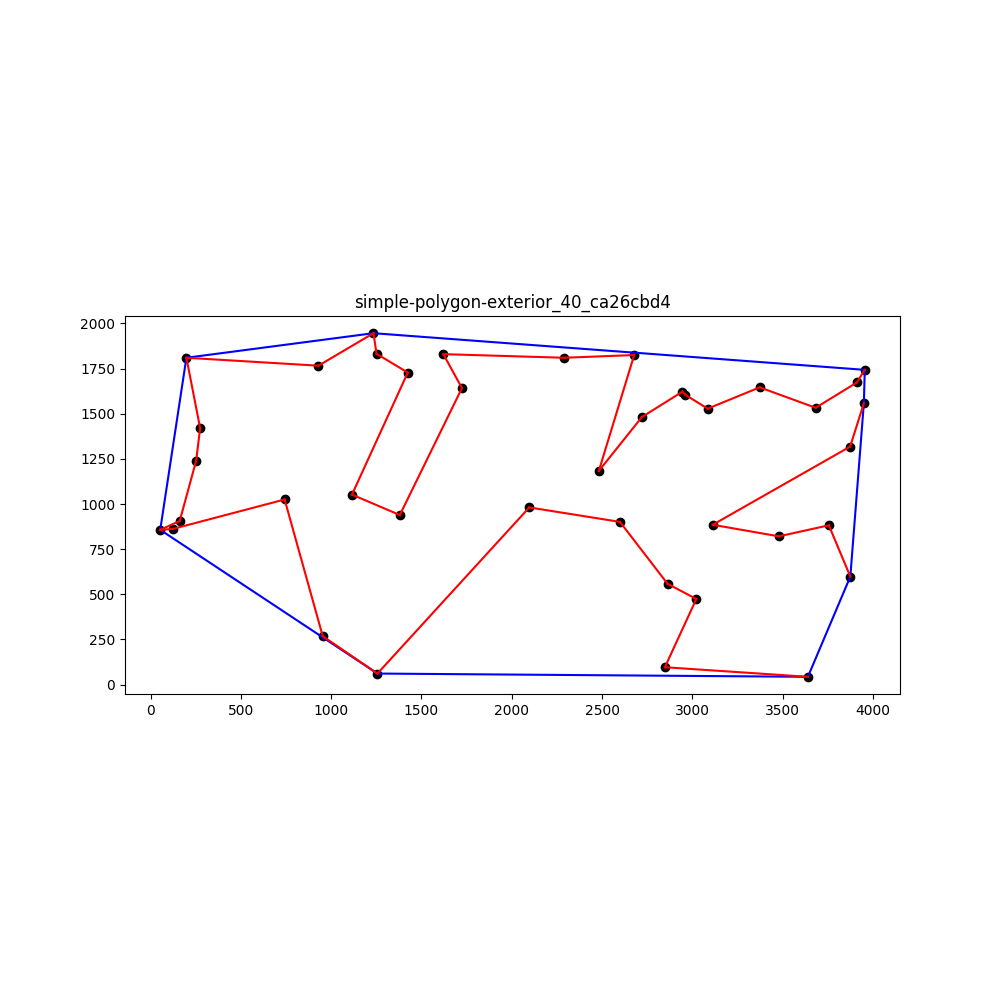}
		\caption{\texttt{simple-polygon-exterior\_40\_ca26cbd4}}
		\label{fig:instances:simple-polygon-exterior}
	\end{subfigure}
	\hfill
	\begin{subfigure}[b]{0.45\textwidth}
		\centering
		\includegraphics[width=\textwidth]{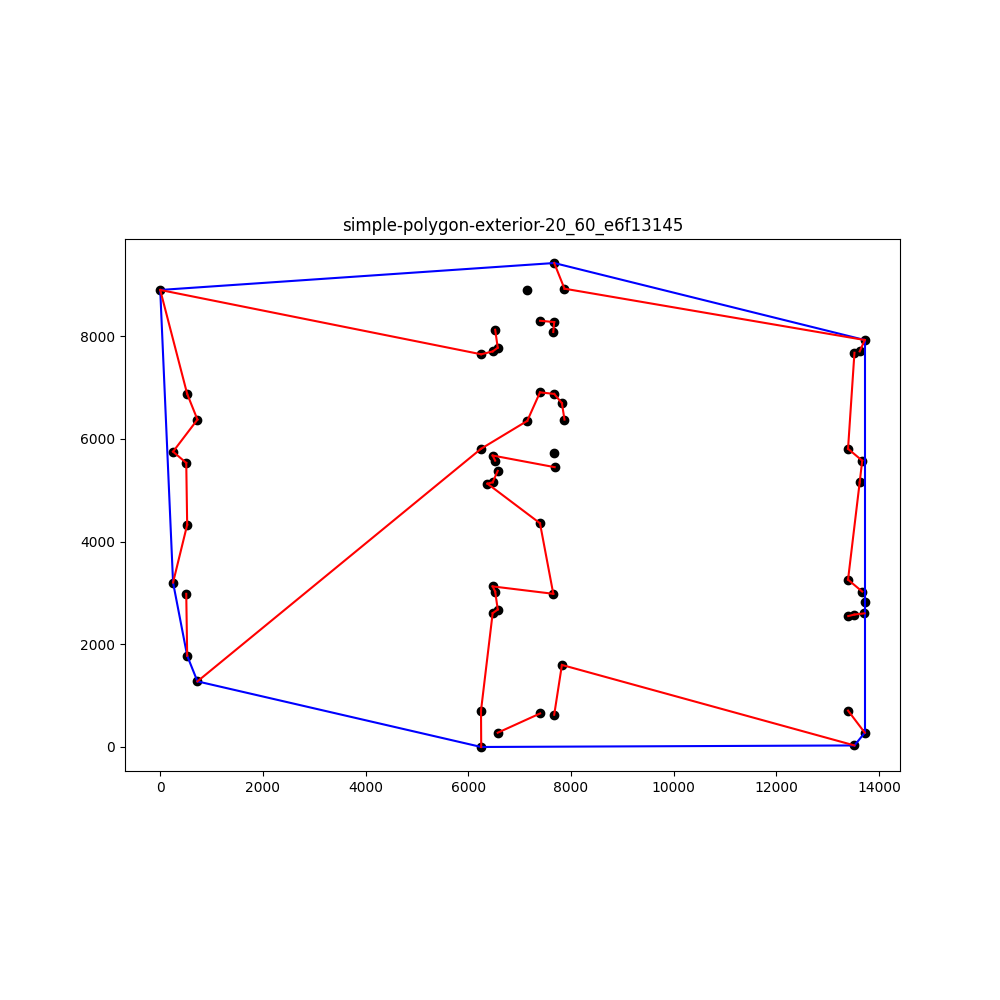}
		\caption{\texttt{simple-polygon-exterior-20\_60\_e6f13145}}
		\label{fig:instances:simple-polygon-exterior-20}
	\end{subfigure}
	\caption{
		Examples of the five instance types used in the challenge. The boundaries are shown in blue, the constrained segments in red, and the points in black.
	}
	\label{fig:instances}
\end{figure}

We developed five distinct instance generators to create a candidate set of instances:
\begin{description}
	\item[ortho] The \emph{ortho} generator produces instances defined by orthogonal polygons specified solely by their boundaries, without additional points or segments.
	It begins with a random rectangle and randomly introduces orthogonal indentations or extensions along the boundary until the desired complexity is achieved. See \cref{fig:instances:ortho} for an example.
	\item[point-set] The \emph{point-set} generator creates instances comprising a set of points in the plane, with the convex hull of these points forming the boundary.
	The points are sampled from structured sets, such as those provided by the TSPLib instances~\cite{reinelt1991tsplib}. See \cref{fig:instances:point-set} for an example.
	\item[simple-polygon] The \emph{simple-polygon} generator produces instances defined by a simple polygon that is given as the boundary without any additional points or segments.
	This generator extends the previous approach by starting with a random tour and applying two-opt swaps to form a simple polygon.
	See \cref{fig:instances:simple-polygon} for an example.
	\item[simple-polygon-exterior] The \emph{simple-polygon-exterior} generator extends the simple-polygon approach by incorporating additional constraints: points and segments define the polygon, while the exterior, bounded by the convex hull of these points, must also be triangulated. Essentially, it augments the previous generator by adding the convex hull.
	See \cref{fig:instances:simple-polygon-exterior} for an example.
	\item[simple-polygon-exterior-20] The \emph{simple-polygon-exterior-20} generator is based on the simple-polygon-exterior generator but randomly deletes 20\% of the segments, such that the polygon no longer partitions the area into simple polygons.
	See \cref{fig:instances:simple-polygon-exterior-20} for an example.
\end{description}

The orthogonal instances were designed to be the easiest to solve, as relatively simple strategies exist for placing the Steiner points.
The point-set instances offer a straightforward approach to iteratively improving the triangulation given their minimal constraints, although effective strategies for placing Steiner points are less apparent.
The most challenging instances were the simple-polygon-exterior-20 cases, which feature unstructured constraints that complicate the application of a unified strategy.
For those, we were not even certain that a solution without obtuse triangles exists.

Given that even solving smaller instances by hand proved extremely difficult, we restricted the instances to a maximum of \num{250} points and generated many small instances, confident that this level of difficulty would be sufficient.
We were also aware that the problem demands a significant amount of exact arithmetic, making it impractical to scale up to larger instances -- especially for student teams.
Overall, \num{150} instances were selected from a larger set of candidates to ensure diverse instances based on various instance features.

\subsection{Categories}

The contest was run in an \emph{Open Class}, in which participants could use any
computing device, any amount of computing time (within the duration of the
contest) and any team composition. 
In the \emph{Junior Class}, a team was required to consist largely of participants who were eligible according to the rules of CG:YRF (the \emph{Young Researchers Forum} of CG Week), defined as not having defended a formal doctorate before 2023. Additionally, a team could include one senior academic advisor, who was allowed to provide guidance and assist with writing the final paper (if the team was selected for inclusion in the proceedings), but all programming tasks had to be performed by the junior team members.

\subsection{Server and Timeline}

The competition was facilitated through a dedicated server at TU Braunschweig,
 accessible at \url{https://cgshop.ibr.cs.tu-bs.de/competition/cg-shop-2025/}.
An initial batch of example instances was made available on July 31, 2024,
followed by the release of the final benchmark set on September 30th, 2024. 
The competition concluded on January 22nd, 2025 (AoE).

Participants were provided with a verification tool as an open-source Python package, available at \url{https://github.com/CG-SHOP/pyutils25}.
This tool provided detailed information on errors and allowed participants to easily investigate and correct their solutions if the server rejected them.
To ensure accurate verification and address potential floating point arithmetic issues, the tool leveraged the CGAL library~\cite{cgal}.
It also provided easily usable Python-bindings to some exact arithmetic operations that we considered potentially useful for the participants such that also users without a C++ background could use them.

\section{Outcomes}

\Cref{tab:team_ranks} shows the results for teams that outperformed the trivial Delaunay Triangulation baseline, ranked by performance.
The first team achieved a near-perfect score of \num{149.813} out of \num{150} and was awarded the first place, while the second team achieved a score of \num{139.769} and was recognized as the best junior team.
These were the only teams to compute solutions without any obtuse triangles across all instances and were invited to contribute to the 2025 SoCG proceedings:

\begin{enumerate}
\item \textbf{Team Naughty NOTers}: Mikkel Abrahamsen, Florestan Brunck, Jacobus Conradi, Benedikt Kolbe and André Nusser~\cite{Challenge2025_1}.
\item \textbf{Team Gwamegi}, also recognized as the best junior team: Taehoon Ahn, Jaegun Lee, Byeonguk Kang and  Hwi Kim~\cite{Challenge2025_2}.
\end{enumerate}


Both winning teams employed local search strategies that iteratively insert, move, and delete Steiner points, which are then completed into a full triangulation via a constrained Delaunay triangulation.
The overall winning team refined these actions with more advanced techniques, including a crossover strategy to merge solutions.

Team scores are detailed in \cref{tab:team_ranks} and further broken down by instance type in \cref{tab:team_scores} and by instance size in \cref{tab:team_scores_instance_size}.

\begin{table}[ht]
    \centering
    \begin{tabular}{lrrrcl}
    \hline
    \textbf{Rank} & \textbf{Team Name} & \textbf{Score} & \textbf{\# feasible} & \textbf{Junior} & \textbf{Comment}\\
    \hline
    1 & Naughty NOTers & 149.813 & 150 & &\\ 
    2 & Gwamegi & 139.769 & 150 & \checkmark &\\ 
    3 & KITriangle & 85.482 & 147 & \checkmark &\\ 
    4 & Obtuse Terminators & 80.756 & 114 & \checkmark &\\ 
    5 & <anonymous> & 61.541 & 81 & \checkmark& \\ 
    6 & die-obtuse & 55.596 & 54 & \checkmark & \\ 
    7 & cheetos & 25.973 & 0 & \checkmark & \\ 
    8 & Delaunay Baseline & 25.935 & 0 & & \emph{Delaunay Triang. as reference} \\ 
    \hline
    \end{tabular}
    \caption{Team rankings with scores and junior team status. Only teams outperforming the baseline are shown. \textbf{\# feasible} denotes the number of instances solved without obtuse triangles.}
    \label{tab:team_ranks}
\end{table}

\begin{table}[ht]
	\centering
	\begin{tabular}{lrrrrrr}
	\hline
	\textbf{Instance Type} & \textbf{ortho} & \textbf{point-set} & \textbf{simple-poly} & \textbf{spoly-ext} & \textbf{spoly-ext-20} \\
	\hline
	Naughty NOTers & 13.000 & 39.955 & 21.955 & 32.904 & 42.000  \\
	Gwamegi & 12.704 & 37.230 & 20.984 & 30.437 & 38.416 \\
	KITriangle & 7.888 & 22.986 & 12.200 & 18.815 & 23.592  \\
	Obtuse Terminators & 8.701 & 23.589 & 13.290 & 15.498 & 19.678 \\
	<anonymous> & 8.628 & 21.855 & 9.613 & 8.679 & 12.765  \\
	die-obtuse & 9.473 & 14.046 & 11.848 & 8.708 & 11.521\\
	cheetos & 2.858 & 7.023 & 4.415 & 5.309 & 6.368  \\
	Delaunay Baseline & 2.909 & 6.946 & 4.404 & 5.309 & 6.368  \\
	\hline
	\end{tabular}
	\caption{Team scores across different instance types. For readability, instance type names have been abbreviated: \textbf{ortho}=\texttt{ortho}, \textbf{point-set}=\texttt{point-set}, \textbf{simple-poly}=\texttt{simple-polygon}, \textbf{spoly-ext}=\texttt{simple-polygon-exterior}, and \textbf{spoly-ext-20}=\texttt{simple-polygon-exterior-20}.}
	\label{tab:team_scores}
\end{table}

\begin{table}[ht]
	\centering
	\begin{tabular}{lrrrrrrrrr}
	\hline
	\textbf{Instance Size} & \textbf{10} & \textbf{20} & \textbf{40} & \textbf{60} & \textbf{80} & \textbf{100} & \textbf{150} & \textbf{250} \\
	\hline
	Naughty NOTers & 24.883 & 24.975 & 19.000 & 22.000 & 13.955 & 17.000 & 14.000 & 14.000  \\
	Gwamegi & 24.509 & 23.989 & 17.874 & 20.341 & 13.096 & 15.553 & 12.376 & 12.031 \\
	KITriangle & 18.100 & 14.652 & 10.342 & 11.826 & 7.519 & 9.171 & 6.766 & 7.107 \\
	Obtuse Terminators & 18.285 & 14.715 & 10.494 & 11.039 & 7.012 & 7.828 & 5.862 & 5.521  \\
	<anonymous> & 16.095 & 13.102 & 8.263 & 8.102 & 6.478 & 3.943 & 3.190 & 2.367 \\
	die-obtuse & 18.802 & 12.807 & 7.581 & 5.552 & 3.501 & 3.190 & 2.052 & 2.112  \\
	cheetos & 10.069 & 7.356 & 3.318 & 2.326 & 1.161 & 1.212 & 0.357 & 0.173  \\
	Delaunay Baseline & 10.057 & 7.336 & 3.318 & 2.322 & 1.159 & 1.212 & 0.357 & 0.173  \\
	\hline
	\end{tabular}
	\caption{Team scores broken down by instance size.}
	\label{tab:team_scores_instance_size}
\end{table}

\begin{figure}
	\centering
	\includegraphics[width=\textwidth]{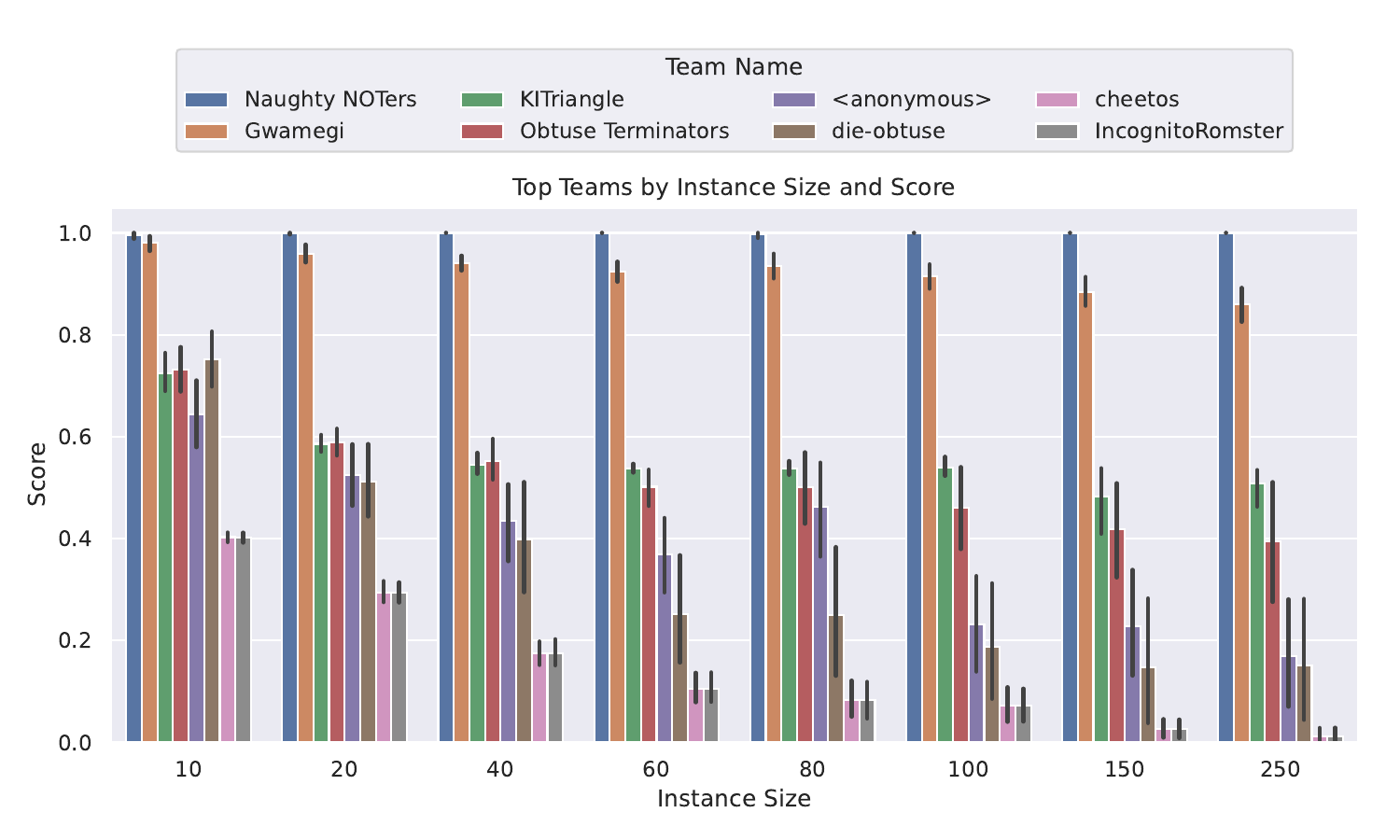}
	\caption{Scores of the top teams across different instance sizes. The x-axis represents the number of items in the instance, and the y-axis shows the corresponding score.}
	\label{fig:top_teams_by_instance_size}
\end{figure}

\begin{figure}
	\centering
	\includegraphics[width=\textwidth]{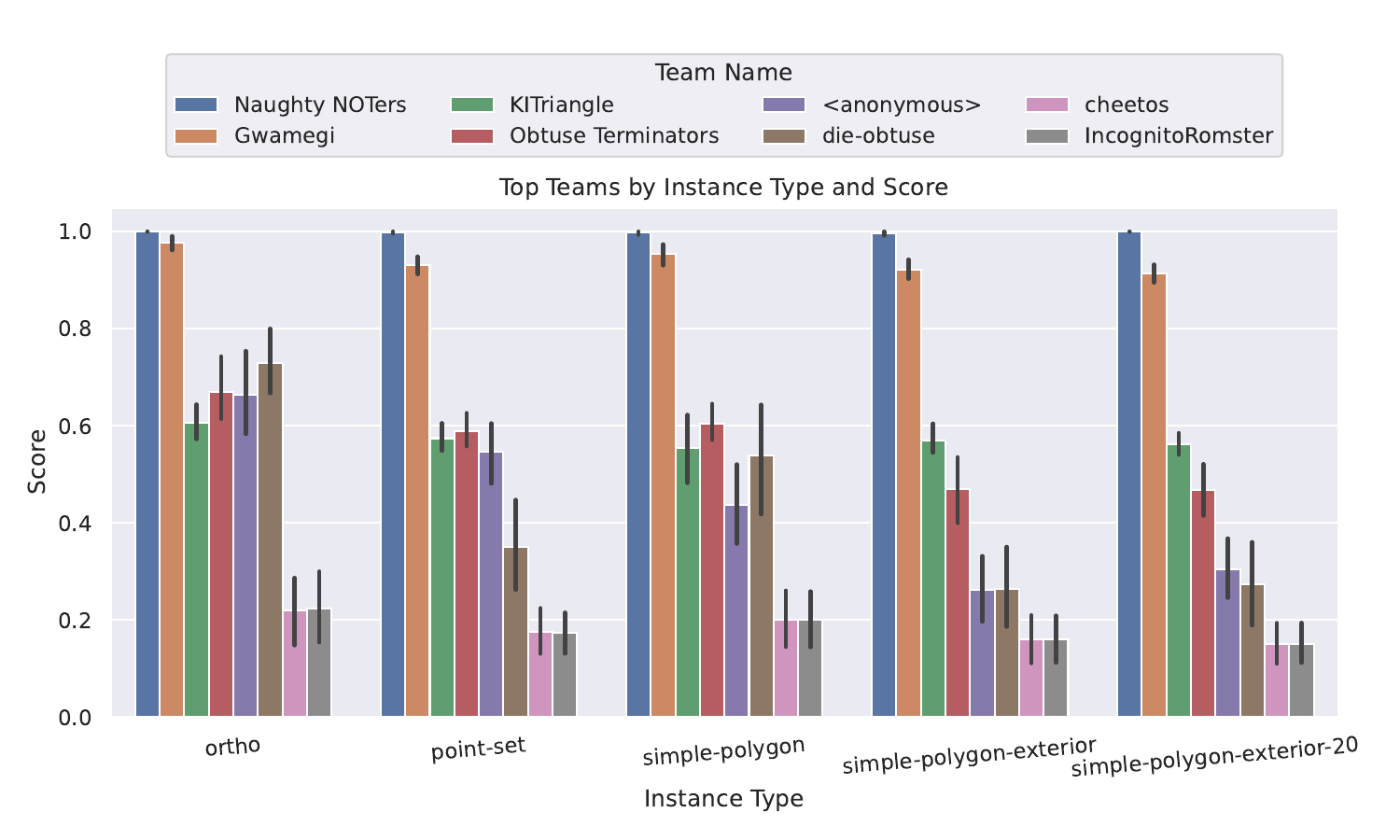}
	\caption{Scores of the top teams across different instance types. The x-axis represents the instance type, and the y-axis shows the corresponding score.}
	\label{fig:top_teams_by_instance_type}
\end{figure}

\Cref{fig:top_teams_by_instance_size} illustrates the performance of the top teams across various instance sizes.
The increasing gap between the top two teams with larger instances suggests that the winning team employed a more effective scaling strategy.
Similarly, \cref{fig:top_teams_by_instance_type} shows performance across instance types.
The \textbf{ortho} instances are confirmed to be the easiest, while the \textbf{simple-polygon-exterior} and \textbf{simple-polygon-exterior-20} instances pose the greatest challenges, with the former appearing slightly more difficult.
Although the fifth team (die-obtuse) ranked third on \textbf{ortho} instances, it could not maintain competitive performance on other types.
The visible variation in rankings between all instance types except \textbf{simple-polygon-exterior} and \textbf{simple-polygon-exterior-20} indicates that the instance types actually demand distinct strategies.

\section{Conclusions}
The 2025 CG:SHOP Challenge motivated a considerable number of teams to engage
in extensive optimization studies. The outcomes promise further insight into
the underlying, important optimization problem. This demonstrates the
importance of considering geometric optimization problems from a practical 
perspective.

\bibliography{bibliography,references}
\end{document}